\documentclass[10pt,a4paper]{book}
\usepackage{udineHyderabad}
\begin{document}

\talktitle{A Quantum Approach to Cosmology, \\
Udine, Septempber 26-29, 2004}
\talkauthors{Antonio Alfonso-Faus
\structure{a}}
\authorstucture[a]{E.U.I.T. Aeronáutica,
Plaza Cardenal Cisneros s/n, 28040 Madrid, Spain}
\shorttitle{A Quantum Approach to Cosmology}
\firstauthor{Antonio Alfonso-Faus}
\begin{abstract}
We present a theory based upon the treatment of the
gravitational field as a sea of gravity quanta,
as defined elsewhere.
The resultant model for the Universe is a static one,
like Einstein first saw, with a new feature:
a local shrinking quantum world that completely
explains the Hubble red shift under a new point of view.
The presently accepted expansion of the Universe is
interpreted here as an apparent effect, as seen from the
Lab system of reference.
The static Universe has immersed in it a local
shrinking atomic world: a fundamental change in
the interpretation of the Hubble's observations.
The conservation principles (momentum, angular
momentum and energy) can be dealt with under 2
different points of view: local (apparent)
and COSMOLOGICAL ("real").
The 2 are in complete agreement with observation.
They are also free of well known contradictions
or paradoxes/incoherencies (i.e. in the Big Bang model).
By dealing now with very well known first principles
(Heisenberg, Mach, de Broglie, Weinberg's relation)
under the same 2 points of view, we arrive at the
conclusion that our new approach is in accordance
with the Einstein's field equations of General Relativity,
and Quantum Mechanics. We consider this to be
a promising first step towards the way of dealing
with the gravitational field coherently both from
the General Relativity and from the Quantum Mechanical
theories. The agreement with the present values of
the cosmological parameters is very satisfactory.
\end{abstract}

\section{Introduction}

The point of view of considering the gravitational
field as a sea of gravity quanta \cite{Alf1}
has been dealt with elsewhere. The published
result there for the mass mg of this quantum is
given by the relation  $m_g =  \hbar /(c^2 t)$,
where t is the age of the Universe
(then today mg is of the order of $2 \times
10^{-66}$ grams).

The first important consequence of the above
approach is the need to introduce a new concept,
that we call the Mass Boom, \cite{Alf2},
\cite{Alf3}, \cite{Alf4} and \cite{Alf5}.
In essence it expresses the property of any
gravitational mass that, due to the emission
of these gravity quanta, having a negative energy,
its mass increases linearly with cosmological time.
This linear dependence
between mass and time makes it possible to
identify the mass of the Universe $M$ with the
cosmological time t. A philosophical statement
like {\em we are made of time} obviously follows and
merits a deep reflexion. Clearly this approach is of
the Machean type.

The second important consequence of this approach
is that the speed of light must decrease with time.
In fact it can be equated to the inverse of t,
$c = 1/t$  \cite{Alf4} and \cite{Alf5}
Then the resultant model
for the Universe is a static one, as Einstein first
proposed, and mathematically stated as $a(t) = ct =$
constant, i.e. a constant cosmological scale factor.
The expansion of the Universe, a generally accepted
interpretation of the red shift, is interpreted here
as an apparent effect seen from the laboratory system
of reference. The reinterpretation of the Hubble red shift
is that the quantum world is shrinking, an effect coming
directly from the time variation of Planck's {\it constant}
\cite{Alf3}, proportional to
$1/t^2$ or equivalently to
$c^2$.

The work we present here is based upon the above
results. We analyze, from this new point of view,
the conservation principles, and solve both: the
Schrödinger equation together with the Einstein
cosmological equations, which represents a first
step in the harmonization of Quantum Mechanics and
Relativity. The conclusion is that the whole approach
is very promising and liberates present theories
(like the Big Bang) from contradictions ands paradoxes.
The agreement with the known numerical values for the
cosmological parameters, as accepted today, is
very satisfactory.

\section{New concepts in the conservation principles}

A summary of the new concepts is as follows:

\begin{itemize}

\item The mass of the gravity quanta,
$m_g =  \hbar /(c^2 t)$.

\item The Mass Boom effect: any gravitational mass $m$
has a time
dependence as $m \ =$ constant $t$
($t$ the age of the Universe).

\item The decrease of the speed of light with time,
$c = 1/t$.

\item The $G = c^3$ relation, following the
Action Principle.

\item Heisenberg and the De Broglie wavelength
(perhaps the Compton wavelength as an alternative),
$\hbar / mc \ =$ constant in the Lab.

\item The $v/c$ constancy as seen from the LAB
system in order to conserve the constancy of the
relativistic relations at any time.

\item The Mass Boom is always present (as long as gravity
is present).

\item The decrease of the speed of light
with time as $c=1/t$,
always present as a consequence of the constancy
of momentum (in the absence of mechanical perturbations).

\item The apparent interpretation of the cosmological
expansion, following the Hubble's observations.

\item The $h = c^2$ relation (which explains the
contraction of the quantum world).

\item Weinberg's relation under a new point
of view: not only explains the quantum of mass at
the local Lab. It explains the Universe as a quantum
black hole whose mass increases linearly with time.

\item The introduction of $H = 1$, the cosmological
Planck's constant, given by the relation
$H = \hbar t^2 = 1$,
which is the essence of the quantum approach to cosmology.

\item The Cosmological Planck's units using $H$,
instead of $\hbar$, defining the cosmological quantum given
by the whole Universe (mass, size and time $M = t$,
and size $ct = 1 = 10^{28} \ cm$
with the constant homogeneous tic given by Planck's time).

\item The fluctuation of the whole Universe,
seen as a quantum black hole (corroborated by
the Weinberg's relation using H).

\item The determination of the age of the Universe
as $t = 10^{61}$ units of time
(the age of the Universe today) and
given by the ratio of Planck's length at
$t = 1$ (the constant length 1028 cm)
and the present value of $10^{-33} \ cm$).

\item The solution to the cosmological Schrödinger
equation coupled with the Einstein's cosmological
equations (harmonization of General Relativity and
Quantum Mechanics).

\item The new entropy concept,
that includes the gravitational entropy:
$S = k M/m_g = M = t$ (for the Universe, \cite{Alf6}).

\end{itemize}

\section{The solution to the Schrödinger cosmological
equation coupled to the Einstein cosmological equations}

The Schrödinger equation can be formulated from a
cosmological point of view by using the "cosmological"
Planck's constant $H = \hbar t^2 = 1$ (a real constant).
The resultant equation is then:

\[
\frac{H^2}{2 m} \ \frac{\partial^2 \Psi}{\partial x^2}
\ + \ V \Psi \ = \ i H \ \frac{\partial \Psi}{\partial t}
\]

We see that all the terms in this equation vary as $1/t$.
Then multiplying by t we have both members of the equation
constant, as in the normal quantum mechanical treatment.
The solution is then, assuming the wave function to be
represented by a product of two functions: one depending
on space and the other depending on time only, in the usual
way one has:

\[
\Psi (x,t) \ = \ const \ (sinx) \ t^2
\]

On the other hand the Einstein cosmological
equations have the solution $a(t) = t^2$ \cite{Alf5}
which coincides with the above time dependence, as seen
from the Lab reference system. Hence we have the same
time dependent solution for both: General Relativity
and Quantum Mechanics.

\section{Conclusions}

The main implications of this work are:
a change in basic paradigms, perhaps the most
important one is the new explanation for gravitation,
in quantum mechanical terms, and coherent with general
relativity. Also a new approach to the entropy concepts,
in particular to the Hawking-Bekenstein treatments,
where the entropy of a black hole is here defined as
linear with mass \cite{Alf7}.
A return to the Einstein
initial cosmological model is the most significant
change in the formulation of cosmological models.

\end{document}